# A Rosetta Stone Relating Conventions In Photo-Meson Partial Wave Analyses

A.M. Sandorfi[a], B. Dey[b], A. Sarantsev[c], L. Tiator[d] and R. Workman[e]

[a]Thomas Jefferson National Accelerator Facility, Newport News, VA 23606, USA
[b]Carnegie Mellon University, Pittsburgh, PA 15213, USA
[c]Helmholtz-Institut für Strahlen- und Kernphysik der Universität Bonn, D-53115 Bonn, Germany
[d]Institut für Kernphysik, Johannes Gutenberg Universität, D-55099 Mainz, Germany
[e]Center for Nuclear Studies, The George Washington University, Washington DC 20052, USA

**Abstract.** A new generation of *complete* experiments in pseudoscalar meson photo-production is being pursued at several laboratories. While new data are emerging, there is some confusion regarding definitions of asymmetries and the conventions used in partial wave analyses (PWA). We present expressions for constructing asymmetries as coordinate-system independent ratios of cross sections, along with the names used for these ratios by different PWA groups. (This update reflects a recent change in the definition of an asymmetry as used by one group.)

The low energy structure of QCD lies encoded in the excited-state spectrum of the nucleon, which is a complicated overlap of many resonances that must be disentangled through partial wave analyses. The spin degrees of freedom in meson photo-production reveal signatures of interfering partial wave strength that are often dramatic and provide essential constraints to PWA solutions.

In single-pseudoscalar meson photoproduction there are 16 possible observables, the cross section ($\sigma$), three asymmetries which to leading order enter the general cross section scaled by a single polarization of either beam, target or recoil ($\Sigma$, $T$, $P$), and three sets of four asymmetries whose leading dependence in the general cross section involves two polarizations of either beam-target ($E$, $G$, $F$, $H$), beam-recoil ($C_{x'}$ $C_{z'}$, $O_{x'}$, $O_{z'}$) or target-recoil ($L_{x'}$, $L_{z'}$, $T_{x'}$, $T_{z'}$). An examination of the literature reveals differences in expressions for these quantities, which arise because the same symbol or *asymmetry name* has been used by different authors to refer to different experimental quantities [1].

The situation is further complicated by the use of several different coordinate systems in the literature, both in PWA papers as well as those reporting experimental results where the construction of a specific detector often naturally lends itself to a particular choice of axes.

Here we construct ratios from the center of mass differential cross sections for detecting a meson with specific orientations of beam, target and recoil polarization, $d\sigma^{B,T,R}(\vec{P}^\gamma, \vec{P}^T, \vec{P}^R)$, and specify these polarization orientations only with vectors constructed from the momenta of the photon ($\vec{p}_\gamma$) and the outgoing meson ($\vec{p}_m$). For this, we form three unit vectors as follows,

$$\hat{p}_1 = \frac{(\vec{p}_\gamma \times \vec{p}_m) \times \vec{p}_\gamma}{\left|(\vec{p}_\gamma \times \vec{p}_m) \times \vec{p}_\gamma\right|}, \qquad \hat{p}_2 = \frac{(\vec{p}_\gamma \times \vec{p}_m)}{\left|(\vec{p}_\gamma \times \vec{p}_m)\right|}, \qquad \hat{p}_3 = \frac{(\vec{p}_\gamma \times \vec{p}_m) \times \vec{p}_m}{\left|(\vec{p}_\gamma \times \vec{p}_m) \times \vec{p}_m\right|}.$$

These can be readily formed from events registered in any detector. Circular beam polarization is designated by the photon helicity as $P_h^\gamma = \pm 1$. Linear beam polarization is designated by the angle ($\phi_\gamma^L$) of the oscillating electric field vector of the ensemble of photons, with positive angles corresponding to a rotation from $\hat{p}_1$ towards $\hat{p}_2$. Unobserved initial polarization states ($s_i$) must be averaged over when constructing cross sections. Similarly, unobserved final polarization states ($s_f$) must be summed.

The following combinations of measurements are used to construct the *single-polarization* ratios.

$$R_S = \frac{\left[d\sigma_1^{B,T,R}\left(\phi_\gamma^L = +\pi/2,\ \text{ave } s_i,\ \text{sum } s_f\right) - d\sigma_2^{B,T,R}\left(\phi_\gamma^L = 0,\ \text{ave } s_i,\ \text{sum } s_f\right)\right]}{\left[d\sigma_1^{B,T,R} + d\sigma_2^{B,T,R}\right]}$$

$$R_T = \frac{\left[d\sigma_1^{B,T,R}\left(\text{ave } s_i,\ \vec{P}^T = +\hat{p}_2,\ \text{sum } s_f\right) - d\sigma_2^{B,T,R}\left(\text{ave } s_i,\ \vec{P}^T = -\hat{p}_2,\ \text{sum } s_f\right)\right]}{\left[d\sigma_1^{B,T,R} + d\sigma_2^{B,T,R}\right]}$$

$$R_P = \frac{\left[d\sigma_1^{B,T,R}\left(\text{ave } s_i,\ \text{ave } s_i,\ \vec{P}^R = +\hat{p}_2\right) - d\sigma_2^{B,T,R}\left(\text{ave } s_i,\ \text{ave } s_i,\ \vec{P}^R = -\hat{p}_2\right)\right]}{\left[d\sigma_1^{B,T,R} + d\sigma_2^{B,T,R}\right]}$$

*Double-polarization* ratios involving polarized beams and targets are constructed as follows.

$$R_E = \frac{\left[d\sigma_1^{B,T,R}\left(P_h^\gamma = +1,\ \vec{P}^T = -\hat{p}_\gamma,\ \text{sum } s_f\right) - d\sigma_2^{B,T,R}\left(P_h^\gamma = +1,\ \vec{P}^T = +\hat{p}_\gamma,\ \text{sum } s_f\right)\right]}{\left[d\sigma_1^{B,T,R} + d\sigma_2^{B,T,R}\right]}$$

$$R_F = \frac{\left[d\sigma_1^{B,T,R}\left(P_h^\gamma = +1,\ \vec{P}^T = +\hat{p}_1,\ \text{sum } s_f\right) - d\sigma_2^{B,T,R}\left(P_h^\gamma = -1,\ \vec{P}^T = +\hat{p}_1,\ \text{sum } s_f\right)\right]}{\left[d\sigma_1^{B,T,R} + d\sigma_2^{B,T,R}\right]}$$

$$R_G = \frac{\left[d\sigma_1^{B,T,R}\left(\phi_\gamma^L = +\pi/4,\ \vec{P}^T = +\hat{p}_\gamma,\ \text{sum } s_f\right) - d\sigma_2^{B,T,R}\left(\phi_\gamma^L = +\pi/4,\ \vec{P}^T = -\hat{p}_\gamma,\ \text{sum } s_f\right)\right]}{\left[d\sigma_1^{B,T,R} + d\sigma_2^{B,T,R}\right]}$$

$$R_H = \frac{\left[d\sigma_1^{B,T,R}\left(\phi_\gamma^L = +\pi/4,\ \vec{P}^T = +\hat{p}_1,\ sum\ s_f\right) - d\sigma_2^{B,T,R}\left(\phi_\gamma^L = +\pi/4,\ \vec{P}^T = -\hat{p}_1,\ sum\ s_f\right)\right]}{\left[d\sigma_1^{B,T,R} + d\sigma_2^{B,T,R}\right]}$$

*Double-polarization* ratios involving the polarization of the beam and the recoil baryon polarization are constructed as follows.

$$R_{C_{x'}} = \frac{\left[d\sigma_1^{B,T,R}\left(P_h^\gamma = +1,\ ave\ s_i,\ \vec{P}^R = +\hat{p}_3\right) - d\sigma_2^{B,T,R}\left(P_h^\gamma = +1,\ ave\ s_i,\ \vec{P}^R = -\hat{p}_3\right)\right]}{\left[d\sigma_1^{B,T,R} + d\sigma_2^{B,T,R}\right]}$$

$$R_{C_{z'}} = \frac{\left[d\sigma_1^{B,T,R}\left(P_h^\gamma = +1,\ ave\ s_i,\ \vec{P}^R = +\hat{p}_m\right) - d\sigma_2^{B,T,R}\left(P_h^\gamma = +1,\ ave\ s_i,\ \vec{P}^R = -\hat{p}_m\right)\right]}{\left[d\sigma_1^{B,T,R} + d\sigma_2^{B,T,R}\right]}$$

$$R_{O_{x'}} = \frac{\left[d\sigma_1^{B,T,R}\left(\phi_\gamma^L = +\pi/4,\ ave\ s_i,\ \vec{P}^R = +\hat{p}_3\right) - d\sigma_2^{B,T,R}\left(\phi_\gamma^L = +\pi/4,\ ave\ s_i,\ \vec{P}^R = -\hat{p}_3\right)\right]}{\left[d\sigma_1^{B,T,R} + d\sigma_2^{B,T,R}\right]}$$

$$R_{O_{z'}} = \frac{\left[d\sigma_1^{B,T,R}\left(\phi_\gamma^L = +\pi/4,\ ave\ s_i,\ \vec{P}^R = +\hat{p}_m\right) - d\sigma_2^{B,T,R}\left(\phi_\gamma^L = +\pi/4,\ ave\ s_i,\ \vec{P}^R = -\hat{p}_m\right)\right]}{\left[d\sigma_1^{B,T,R} + d\sigma_2^{B,T,R}\right]}$$

*Double-polarization* ratios involving the polarization of the target and recoil baryon polarization are constructed as follows.

$$R_{L_{x'}} = \frac{\left[d\sigma_1^{B,T,R}\left(ave\ s_i,\ \vec{P}^T = +\hat{p}_\gamma,\ \vec{P}^R = +\hat{p}_3\right) - d\sigma_2^{B,T,R}\left(ave\ s_i,\ \vec{P}^T = +\hat{p}_\gamma,\ \vec{P}^R = -\hat{p}_3\right)\right]}{\left[d\sigma_1^{B,T,R} + d\sigma_2^{B,T,R}\right]}$$

$$R_{L_{z'}} = \frac{\left[d\sigma_1^{B,T,R}\left(ave\ s_i,\ \vec{P}^T = +\hat{p}_\gamma,\ \vec{P}^R = +\hat{p}_m\right) - d\sigma_2^{B,T,R}\left(ave\ s_i,\ \vec{P}^T = +\hat{p}_\gamma,\ \vec{P}^R = -\hat{p}_m\right)\right]}{\left[d\sigma_1^{B,T,R} + d\sigma_2^{B,T,R}\right]}$$

$$R_{T_{x'}} = \frac{\left[d\sigma_1^{B,T,R}\left(ave\ s_i,\ \vec{P}^T = +\hat{p}_1,\ \vec{P}^R = +\hat{p}_3\right) - d\sigma_2^{B,T,R}\left(ave\ s_i,\ \vec{P}^T = +\hat{p}_1,\ \vec{P}^R = -\hat{p}_3\right)\right]}{\left[d\sigma_1^{B,T,R} + d\sigma_2^{B,T,R}\right]}$$

$$R_{T_{z'}} = \frac{\left[d\sigma_1^{B,T,R}\left(ave\ s_i,\ \vec{P}^T = +\hat{p}_1,\ \vec{P}^R = +\hat{p}_m\right) - d\sigma_2^{B,T,R}\left(ave\ s_i,\ \vec{P}^T = +\hat{p}_1,\ \vec{P}^R = -\hat{p}_m\right)\right]}{\left[d\sigma_1^{B,T,R} + d\sigma_2^{B,T,R}\right]}$$

The names used for these ratios by different groups carrying out partial wave analyses are listed in Table 1. In all cases, the magnitudes are identical but for many ratios the signs vary. To ensure data are correctly incorporated into partial wave analyses, we encourage experimental groups to specify their sign convention relative to Table 1 when reporting new experimental results. (Note added in this update: *Table 1 reflects a recent change by the BoGa PWA in the sign of their definition of E.* )

**TABLE 1.** Names for the coordinate-independent polarization ratios as used in the partial wave analyses of the groups from MAID [2], SAID [3], Bonn-Gatchina (BoGa) [4], Carnegie Mellon (CMU) [5] and JLab-EBAC [1].

|            | MAID, SAID | CMU       | SHKL-EBAC, BoGa |
|------------|------------|-----------|-----------------|
| $R_S$      | Σ          | Σ         | Σ               |
| $R_T$      | T          | T         | T               |
| $R_P$      | P          | P         | P               |
| $R_E$      | E          | –E        | E               |
| $R_F$      | F          | F         | F               |
| $R_G$      | G          | G         | G               |
| $R_H$      | –H         | H         | H               |
| $R_{Cx'}$  | $-C_{x'}$  | $C_{x'}$  | $C_{x'}$        |
| $R_{Cz'}$  | $-C_{z'}$  | $C_{z'}$  | $C_{z'}$        |
| $R_{Ox'}$  | $-O_{x'}$  | $O_{x'}$  | $O_{x'}$        |
| $R_{Oz'}$  | $-O_{z'}$  | $O_{z'}$  | $O_{z'}$        |
| $R_{Lx'}$  | $-L_{x'}$  | $L_{x'}$  | $L_{x'}$        |
| $R_{Lz'}$  | $L_{z'}$   | $L_{z'}$  | $L_{z'}$        |
| $R_{Tx'}$  | $T_{x'}$   | $T_{x'}$  | $T_{x'}$        |
| $R_{Tz'}$  | $T_{z'}$   | $T_{z'}$  | $T_{z'}$        |

## ACKNOWLEDGMENTS

This work was supported by the US Department of Energy, Office of Nuclear Physics Division, under contract no. DE-AC05-06OR23177 under which Jefferson Science Associates operate Jefferson Laboratory, and also by US Department of Energy Grants DE-FG02-87ER40315 and DE-FG02-99ER41110. Support has also been provided by the Deutsche Forschungsgemeinschaft (DFG).